\documentclass[runningheads]{llncs}

\usepackage{graphicx}
\usepackage{multirow}
\usepackage{subcaption}
\usepackage{algorithm}
\usepackage{algpseudocode}
\usepackage{amsmath}
\usepackage{amssymb}
\usepackage{newtxmath}
\usepackage{url}
\usepackage{booktabs}
\usepackage{hyperref}
\usepackage[symbol*]{footmisc}

\newcommand{\numfootnote}[2]{%
    \begingroup
    \renewcommand{\thefootnote}{#1}%
    \footnote{#2}%
    \endgroup
    \setcounter{footnote}{#1}%
    \addtocounter{footnote}{-1}%
}

\begin{document}
\title{VM-Rec: A Variational Mapping Approach for Cold-start User Recommendation}
\titlerunning{VM-Rec}
%
%

\author{Linan Zheng\inst{1,2} \and
Jiale Chen\inst{1,2} \and
Pengsheng Liu\inst{3} \and
Guangfa Zhang\inst{1,2} \and
Jinyun Fang\inst{2}\thanks{Corresponding author.}}

\authorrunning{L. Zheng et al.}

\institute{
University of Chinese Academy of Sciences, Beijing, China
\email{zhenglinan21@mails.ucas.ac.cn}
\and
Institute of Computing Technology, Chinese Academy of Sciences, Beijing, China \email{chenjiale21s@ict.ac.cn, zgf523@gmail.com, fangjy@ict.ac.cn}
\and
College of Big Data and Information Engineering, Guizhou University, Guiyang, China 
\email{ie.psliu21@gzu.edu.cn}
}

%
\maketitle              
\begin{abstract}
The cold-start problem is a common challenge for most recommender systems. The practical application of most cold-start methods is hindered by the deficiency in auxiliary content information for users. Furthermore, most methods often require simultaneous updates to extensive parameters of recommender models, resulting in high training costs, especially in large-scale industrial scenarios. We observe that the model can generate expressive embeddings for warm users with relatively more interactions. Initially, these users were cold-start users, and after transitioning to warm users, they exhibit clustering patterns in their embeddings with consistent initial interactions. Motivated by this, we propose a \textbf{V}ariational \textbf{M}apping approach for cold-start user \textbf{Rec}ommendation (\textbf{VM-Rec}), mapping from few initial interactions to expressive embeddings for cold-start users. Specifically, we encode the initial interactions into a latent representation, where each dimension disentangledly signifies the degree of association with each warm user. Subsequently, we utilize this latent representation as the parameters for the mapping function, mapping (decoding) it into an expressive embedding, which can be integrated into a pre-trained recommender model directly. Our method is evaluated on three datasets, demonstrating superior performance compared to other popular cold-start methods.\footnote[2]{Code is available at \url{https://github.com/Linan2018/VM-Rec}.}
\let\thefootnote\relax\footnotetext{APWeb-WAIM 2024, LNCS 14962, pp. 209–224, 2024. \url{https://doi.org/10.1007/978-981-97-7235-3_14}}

\end{abstract}

\section{Introduction}\label{Introduction}
Recommender systems have been widely applied in many fields nowadays, providing personalized item lists to users. Typically, recommender systems model user interaction information into the user's ID embedding, which is highly beneficial for personalization\cite{guo2016entity,zhao2018learning}. However, models struggle to form embeddings that sufficiently reflect user preferences when there are insufficient interactions\cite{yuan2020parameter,yuan2023go}.

In recent years, the cold-start  problem has received widespread attention. Many models rely on auxiliary information\cite{dropoutnet,zhu2020recommendation}, which is easily obtained on the item side due to the rich descriptions or modal information associated with items. However, obtaining high-quality profiles (such as age, gender, and occupation) of users poses a challenge for privacy concerns, and makes it difficult to apply in practice for cold-start user recommendation.
In addition, many models necessitate the joint updating of parameters in the original recommender model. The deployment cost of such models in large-scale industrial recommender systems is relatively high, and altering the parameters of the recommender model is prone to lead to a seesaw phenomenon\cite{chen2022generative}, thereby affecting the effectiveness of recommendations.

To address the aforementioned issues, our approach is built upon the following two concepts: 1) It is a remarkable fact that a considerable number of existing users were also cold-start users initially, and we can obtain both their initial interactions and sufficiently expressive embeddings through a well-trained general recommender model. 2) We observed that when existing users' initial interactions are consistent, the resulting learned embeddings exhibit similarity or clustering characteristics, 
\begin{figure}[t]
  \centering
  \begin{subfigure}{0.38\columnwidth}
    \centering
    \includegraphics[width=0.898\linewidth]{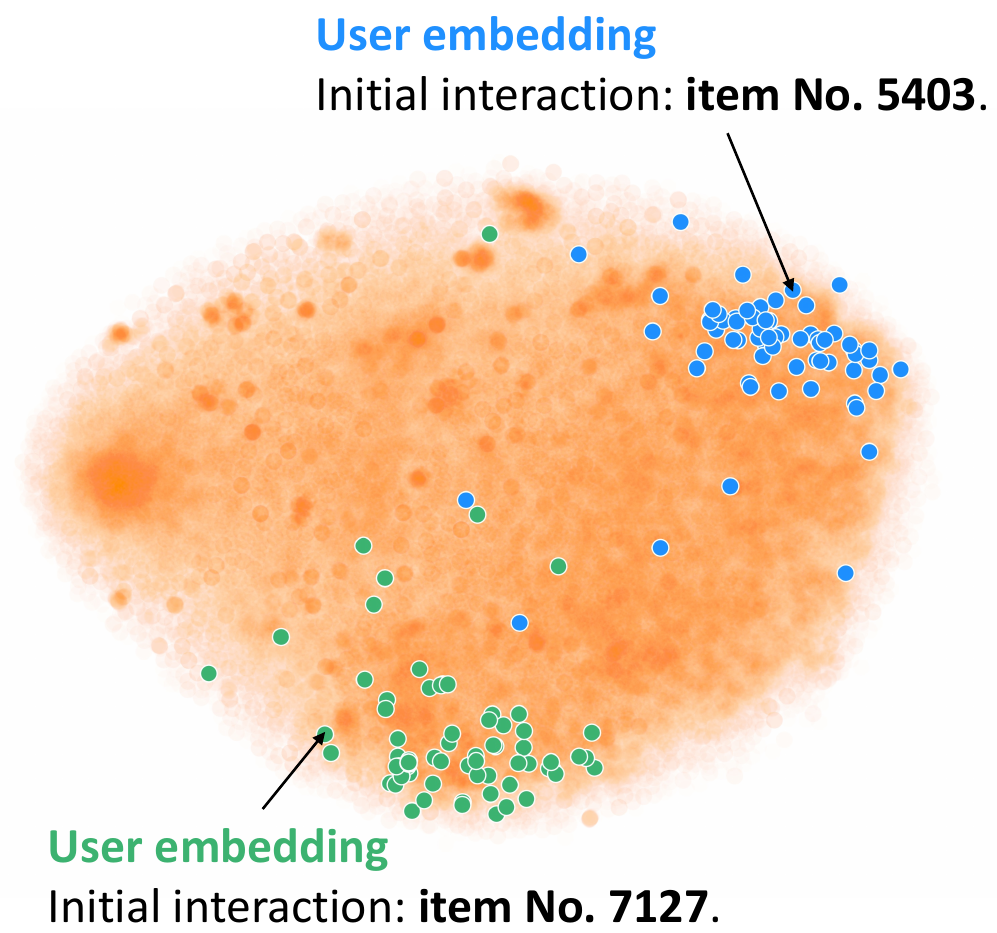}
    \caption{Visualization of pre-trained warm user embeddings using T-SNE\cite{van2008visualizing}.}
    \label{fig:a}
  \end{subfigure}\hfill
  \hspace{1em}
  \begin{subfigure}{0.58\columnwidth}
    \centering
    \includegraphics[width=0.75\linewidth]{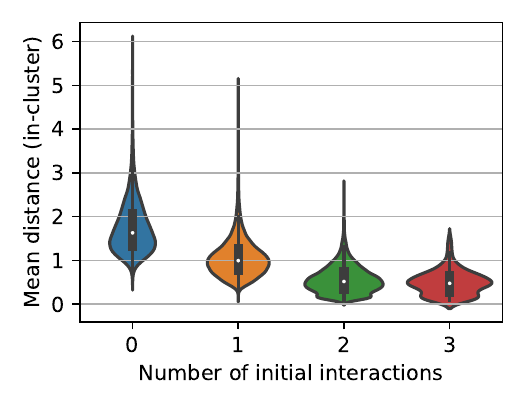}
    \caption{The mean distances (in-cluster) corresponding to varying numbers of initial interactions from 0 to 3 are 1.74, 1.05, 0.56, and 0.49, respectively.}
    \label{fig:b}
  \end{subfigure}
  \vspace{-0.1cm}
  \caption{Exploration of the distribution of embeddings using the BPR model on the Ali Display Ad Click dataset.
  }
  \vspace{-0.4cm}
  \label{fig:subfigures}
\end{figure}
as illustrated in Figure \ref{fig:a}. Furthermore, by calculating the average distance between an embedding and other embeddings within its designated cluster (where users with the same initial interactions are grouped into clusters, each requiring a minimum of two distinct embeddings), we observe that a greater number of initial interactions, indicative of finer cluster divisions, results in reduced average distances, as illustrated in Figure \ref{fig:b}.  When the
number of initial interactions is 0, all embeddings belong to the same cluster; however, with the initiation of cluster formation, a significant reduction in the average distance among the embeddings becomes evident. This finding aligns with our intuitive expectations, and thus, we can conclude that users' initial interactions to some extent reveal the approximate distribution of their embeddings.
Hence, we can align the initial interactions and embeddings by utilizing the initial interactions to predict embeddings that reflect stable interests.

To achieve this goal, we propose a personalized \textbf{V}ariational \textbf{M}apping approach for user cold-start \textbf{Rec}ommendation (\textbf{VM-Rec}), which generates a personalized mapping function based on a cold-start user's initial interactions to predict the embedding.
We recognize that there are two difficulties in modeling a personalized mapping function: 1) it is difficult to characterize the parameter distribution of the mapping function; and 2) the mapping function is susceptible to noise when there are few initial interactions to reflect the user’s interests\cite{velf}. 
To address these issues, we introduce a  variational distribution that closely approximates the true parameter distribution to generate the parameters of the personalized mapping function, rather than directly learning the parameter values. 
Furthermore, given the vast number of existing users, not all embeddings prove efficacious for specific cold-start user. For the sake of computational efficiency and interpretability, the mapping function is modeled as a sparse linear model, where each parameter characterizes the degree of association between each existing warm user and the cold-start user. 
Each embedding of an existing warm user is treated as a separate input feature, and the mapping function is used to predict the embedding of the cold-start user. The predicted embeddings are not generated through optimization steps, which circumvents the issue where using limited interaction samples fails to update initial user embeddings into embeddings with strong expressivity. 

The main contributions of this work are summarized as follows:

1. We introduce a novel approach, VM-Rec, designed to generate expressive cold-start user embeddings without the need for auxiliary content information.

2. VM-Rec is applicable to off-the-shelf recommendation, adaptable to various recommender models, and does not require fine-tuning the base model.

3. We evaluate VM-Rec based on different recommender models and experimental results demonstrate its effectiveness in alleviating the user cold-start problem compared to other approaches on three real-world datasets.


\section{Related Works}



In recent years, methods for addressing cold-start  issues can be broadly categorized into two types: content-based and meta-learning-based. 
The fundamental idea of content-based methods involves processing user/item content features through specific models to generate warm embeddings.
For instance, \cite{dropoutnet,mwuf} utilizes the average of content embeddings or transformed content features to serve as or warm up the cold embeddings. In \cite{ouyang2021learning,metacf,hao2021pre,lu2020meta,liu2023Boosting}, structured information are utilized to generate embeddings. Meanwhile, \cite{wei2021contrastive,zhou2023contrastive} leverage a contrastive learning framework to investigate the correlations between content features and collaborative signals. In \cite{chen2022generative}, generative adversarial networks (GAN)\cite{gan} are introduced to ensure that the generated embeddings align more closely with the distribution of warm embeddings.  \cite{zhao2022improving,velf} investigate the latent variable distribution over side information to generate ID embeddings, and \cite{wdof} utilizes latent variable distribution to model user preferences. 
Meta-learning-based methods introduce the concept of Model-Agnostic Meta-Learning (MAML)\cite{finn2017model} into recommender systems to tackle user cold-start issues. 
For instance, \cite{lee2019melu,metaemb,kim2023meta} utilize meta-learning to generate desirable initial embeddings. \cite{metacf,hao2021pre,lu2020meta,liu2023Boosting} utilize structured information from user-item interaction graphs to generate embeddings for new nodes.

Generally, content-based methods heavily rely on high-quality side information, while meta-learning-based methods are prone to overlooking the differences in user behavior distribution due to varying degrees of familiarity with the system and the majority will dominate during the adaptation phase, which hampers the personalization of cold-start user recommendation\cite{dai2021poso}. Additionally, a sophisticated learning rate needs to be designed to update the initial embedding to an expressive embeddings with limited samples\cite{metacf}. However, VM-Rec employs embedding mapping, which is not dependent on side information and does not require updating embeddings through optimization steps. Furthermore, VM-Rec can adapt to different recommender models without fine-tuning model parameters.

\section{Problem Definition}

Let $\mathcal{U}=\mathcal{U}^{e} \cup \mathcal{U}^{c}$ denote the set of all users, where $\mathcal{U}^{e}=\{u^{e}_1,  \dots, u^{e}_{|\mathcal{U}^{e}|}\}$ denotes the set of existing users, and $\mathcal{U}^c=\{u_1^c,  \ldots, u_{|\mathcal{U}^c|}^c\}$ denotes the set of cold-start users, with $\mathcal{U}^{e}$ and $\mathcal{U}^{c}$ being disjoint. Let $\mathcal{V}=\{v_1, \dots, v_{|\mathcal{V}|}\}$ represent the set of all items, and let $I(u)\subseteq \mathcal{V}$ denote the set of items that user $u$ has interacted with. The pre-trained recommender model is represented by $r(\cdot;\Phi^U, \Phi^V, \Theta_{\text{RS}})$, where $\Phi^U$ with dimension $|\mathcal{U}^{e}|\times d$ denotes the embeddings of all existing users $\mathcal{U}^{e}$, and for each $u^e \in \mathcal{U}^{e}$, the corresponding embedding $\phi_{u^e}$ can be obtained. However, for each $u^c \in \mathcal{U}^{c}$, the corresponding embedding of $u^c$ is missing, as the recommender model is not updated in real-time and embeddings are not computed in a timely manner for cold-start users. $\Phi^V$ with dimension $|\mathcal{V}|\times d$ denotes the embeddings of all items, and for each $v \in V$, its corresponding embedding is $\phi_v$.
$\Theta_{\text{RS}}$ denotes other parameters of the recommender model, such as the weights of multi-layer perceptrons.

For the recommender model $r$, given a user-item pair $(u, v)$, the model can calculate the estimated click-through rate or ranking score of user $u$ on item $i$ as $r(u, v;\Phi^U, \Phi^V, \Theta_{\text{RS}})$. Due to the missing embeddings of cold-start users, the use of randomly initialized embeddings $\phi_{\text{rand}}$ can result in inaccurate recommendations. Therefore, we aim to predict an expressive user embedding $\hat{\phi}_{u^c}$ for the cold-start user $u^c$ that can be effectively integrated into the pre-trained recommender model $r$ to achieve more accurate results.

\section{Methodology}

\begin{figure*}[t]
  \centering
  \includegraphics[width=\linewidth]{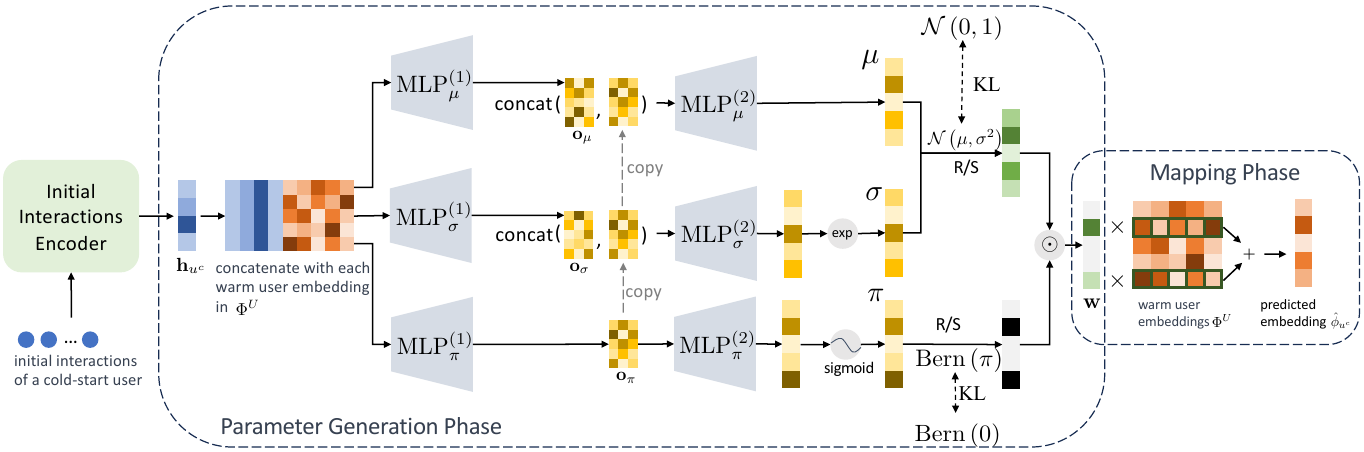}
  \vspace{-0.1cm}
  \caption{
  Predicting an expressive embedding for cold-start users involves two main stages:
\
{\textbf{1. Parameter Generation Phase}}: Based on the encoding of initial interactions, a variational spike-and-slab distribution is generated, from which parameters $\mathbf{w}$ for personalized mapping functions are through the process of reparameterization or sampling (R/S).
\textbf{2. Mapping Phase}: The mapping phase effectively involves selecting specific pre-trained warm user embeddings $\Phi^U$ for weighted summation, and obtains the predicted embedding $\hat{\phi}_{u^c}$.
  }
  \vspace{-0.4cm}

  \label{fig-model}
\end{figure*}

In this paper, we propose a variational personalized mapping approach, called VM-Rec, to generate expressive embeddings for cold-start users. The overall process is illustrated in Figure \ref{fig-model}. 

For a cold-start user $u^c \in \mathcal{U}^c$, a personalized mapping function $f(\cdot;\mathbf{w})$ is generated. The mapping function parameters $\mathbf{w}$ are generated by
the parameter generator $g$ with parameter $\Theta_{\text{PG}}$. $g$ takes the initial interactions $I(u^c)$ (typically, the number of items interacted by a cold-start user is relatively small) as input and outputs the parameters of the personalized mapping function, as $\mathbf{w}= g(I(u^c);\Theta_{\text{PG}})$.
Subsequently, an expressive user embedding $\hat{\phi}_{u^c}=f(\Phi^U;\mathbf{w})$ is predicted for the cold-start user to provide the recommender model with the ability to calculate relatively accurate click-through rates or ranking scores. 



\subsection{Personalized Mapping Approach}
\subsubsection{Initial Interactions Encoder}

In order to generate a personalized mapping function given limited initial interactions $I(u^c) = \{ v_1^{\text{init}}, \dots, v_K^{\text{init}}\}$of a cold-start user $u^c$, with a length of $K$, we first obtain the item embeddings for each item in the initial interaction sequence by utilizing the item embedding layer of a pre-trained recommender model. We then stack $K$ embeddings to obtain $\mathbf{X}_{u^c} = (\phi_{v_1^{\text{init}}}, \dots, \phi_{v_K^{\text{init}}})^{\top}$ with dimension $K\times d$.
Considering the varying number of initial interactions of cold-start users and the potentially different impact of interacted items, we utilize a self-attention layer \cite{vaswani2017attention} to effectively encode the interaction sequence into a tensor representation $\mathbf{h}_{u^c}$ with dimension $d$. This representation, $\mathbf{h}_{u^c}$, will be used to generate the parameters of the personalized mapping function in subsequent steps.


\vspace{-0.1cm}
\subsubsection{Variational Spike-and-slab Distribution}

Given the large number of existing(warm) users, not all warm users' embeddings are significantly associated with the initial interactions of a specific cold-start user. As a result, the predicted embedding of the cold-start user is based only on a few embeddings of warm users.
 As discussed in Section \ref{Introduction}, we can describe the personalized mapping function as a sparse linear model, where most parameters are zero.  
Specifically, we adopt a variational approach and model the distribution of each parameter as a spike-and-slab distribution $q(\mathbf{w}\mid I(u^c))$, which serves as an approximation to the true posterior distribution $p(\mathbf{w}\mid I(u^c))$. The spike-and-slab distribution\cite{sas} is a Bayesian framework-based distribution that facilitates obtaining a sparse model. Sampling from the spike-and-slab distribution can be viewed as a two-stage process: firstly, a random variable $s_i \sim \operatorname{Bernoulli}\left(\pi_i\right)$ is sampled 
from the Bernoulli distribution; then, if $s_i=1$,  $w_i$ is sampled from the Gaussian distribution, otherwise, $w_i$ attains the fixed value $0$, which can be expressed as:
\begin{equation}
\label{eq-sas}
\begin{array}{rll}
w_i \mid s_i=1  &\sim \mathcal{N}\left(\mu_i, \sigma_i^2\right), \\
w_i \mid s_i=0  &\sim \delta\left(w_i\right).
\end{array}
\end{equation}
where $i \in [1,\left|\mathcal{U}^e\right| ]$ as each parameter $w_i$ corresponds to an existing user $u_i^e$ and characterizes the degree of association with that user. From the Equation \ref{eq-sas}, we can observe that the parameters of the Bernoulli distribution control the sparsity of the model parameters, while the parameters of the Gaussian distribution control the both mean and variance of the non-zero parameters in the model. 

For cold-start users with different initial interactions, we can finely tune the probability of each parameter being \textit{sparse} by adjusting the parameters of the Bernoulli distribution in the spike-and-slab distribution. Through the learning process, we selectively mask the embeddings of existing users that are irrelevant to the cold-start user. In Section \ref{Ablation}, we will further compare the performance of different variational distributions used for generating sparse models.

\vspace{-0.12cm}
\subsubsection{Parameter Generation Phase}


To facilitate gradient backpropagation, we employ a reparameterization trick to obtain the parameters $\mathbf{w}$, which allows us to optimize the parameters of earlier layers through gradient descent. 
As the Bernoulli distribution is a discrete distribution, we utilize the Gumbel Softmax trick\cite{gumbel-softmax}, a widely used technique for reparameterizing discrete random variables as continuous ones for efficient gradient-based optimization, to reparameterize the random variable $s_i$ as $\tilde{s}_{i}=\operatorname{Gumbel-Softmax}(\pi_i;g_i)$, where $g_i \sim \operatorname{Gumbel}\left(0, 1\right)$.
After obtaining the reparameterized $\tilde{s}_{i}$, we can calculate each reparameterized weight $\tilde{w}_i$ of the personalized mapping function as follows\cite{kingma2013auto}:

\begin{equation}
\label{repara}
\begin{array}{lll}
\tilde{w}_i \mid \tilde{s}_i &=\tilde{s}_{i} \cdot \left(\mu_{i} + \epsilon_{i} \sigma_{i}\right) + 0 \cdot \left(1-\tilde{s}_{i} \right)  =\tilde{s}_{i} \cdot \left(\mu_{i} + \epsilon_{i} \sigma_{i}\right) 
\end{array}
\end{equation}
where $\epsilon_{i} \sim \mathcal{N}\left(0, 1\right)$. 
By introducing two types of noise variables, namely $g_i$ and $\epsilon_i$, independent of $\pi_i$, $\mu_i$, and $\sigma_i$, we can obtain a reparameterized $\tilde{w}_i$. This allows for the computation of gradients and updating model parameters.

Essentially, our objective is to utilize the spike-and-slab distribution to create personalized mappings akin to a sparse attention mechanism acting upon the entire set of warm user embeddings. The parameters $\pi_i$ of the distribution determine the \textit{selection} of a warm user embedding, while $\mu_i$ and $\sigma_i$ are the factors directly responsible for determining the specific values of the attention weights. 

In order to compute the parameters $\pi_i$, $\mu_i$, and $\sigma_i$ of the spike-and-slab distribution, we commence by concatenating $\mathbf{h}_{u^c}$ and $\phi_{u^e_i}$ to form the input of the neural networks $\operatorname{MLP}^{(1)}_{\pi/\mu/\sigma}$, resulting in $\mathbf{o}_{\pi_i}$, $\mathbf{o}_{\mu_i}$, and $\mathbf{o}_{\sigma_i}$.
Furthermore, to enhance the network's ability to learn $\mu_i$ and $\sigma_i$ more effectively, the information from $\mathbf{o}_{\pi_i}$ is shared within the network responsible for generating $\mu_i$ and $\sigma_i$. By employing an appropriate activation function, we deduce the parameters $\pi_i$, $\mu_i$, and $\sigma_i$ of the spike-and-slab distribution, as illustrated in Figure \ref{fig-model}.







Thus, we are able to compute $\tilde{w}_i$ using Equation \ref{repara}.  The final parameters of the personalized mapping function are obtained by applying masked Softmax, which applies to the selected non-zero weights to maintain parameter sparsity, 
thereby excluding the unselected weights from computation.


\begin{equation}
\label{softmax}
w_i=\operatorname{Masked-Softmax} (\tilde{\mathbf{w} })_i
= \frac 
{\vmathbb{1}\left(\mathcal{C}\right)e^{\tilde{w}_i}}
{\sum_{j=1}^{|\mathcal{U}^{e}|} \vmathbb{1}\left(\mathcal{C}\right)e^{\tilde{w}_i}} 
\end{equation}
where $\vmathbb{1}(\mathcal{C})$ is an indicator function that takes the value $1$ if condition $\mathcal{C}$ is true, and $0$ otherwise. 
During the training process, as we employ both the initial interactions of warm users and their embeddings as ground truth, it is essential to apply masking to the embedding corresponding to the user themselves. This precaution prevents trivial solutions, where 
only the self-embedding is chosen while others are not selected, akin to a leakage of warm embedding.
However, during the inference phase, such masking is unnecessary because we lack knowledge about whether the trained embedding table includes the ID of the cold-start user. Therefore, the masking conditions are as follows:
\vspace{-0.06cm}
\begin{equation}
\begin{aligned}
& \mathcal{C}_\text{train} = \left\{\tilde{w}_i\ne0 \quad \text{or} \quad
i \ne \operatorname{idx}\left(u^c\right)\right\}\\
& \mathcal{C}_\text{infer} = \left\{ \tilde{w}_i\ne0 \right\} 
\end{aligned}
\end{equation}
\vspace{-0.07cm}
where $\operatorname{idx}\left(u^c\right)$ denotes the retrieval of the index of the user $u^c$ within the embedding table. 

\vspace{-0.12cm}
\subsubsection{Mapping Phase}

After obtaining the weights of the personalized mapping function $f(\cdot;\mathbf{w})$, we can use the embeddings of existing users to predict the embedding of a cold-start user $u_c$, as $\hat{\phi}_{u^c}=\sum_{i=0}^{\left|\mathcal{U}^e\right|}w_i\phi _{u^e_i}$. 
where $\phi_{u^e_i}$ represents the pre-trained embedding of an existing user $u^e_i$, while $w_i$ reflects the contribution of $u^e_i$ towards predicting the embedding of the cold-start user $u_c$, and $w_i$ can be considered as an attention weight assigned to the existing embeddings, allowing for the computation of the predicted embedding $\hat{\phi}_{u^c}$ for $u_c$.


\vspace{-0.05cm}
\subsection{ Complexity Analysis }

For a cold-start user, during the parameter generation phase, we initially encode K interactions with a single-layer self-attention layer. The computational complexity of this step is $O(K^2d)$. Subsequently, in the generation of the parameters for the variational distribution, as illustrated in Figure \ref{fig-model} (both employing single-layer MLP) and Equation \ref{softmax}, the computational complexity of this step can be expressed as $O(3 \cdot N \cdot 2d \cdot d' + (N \cdot d' \cdot 1 + N \cdot 2d' \cdot 1 + N \cdot 2d' \cdot 1) + N) = O(Ndd')$, where $d'$ represents the number of neurons in the hidden layer of the MLP, and $N$ denotes the quantity of warm embeddings used (typically, $N = |\mathcal{U}^{e}|$). The computational complexity during the mapping phase is $O(Nd)$. 
Therefore, the overall computational complexity is $O(K^2d + Ndd' + Nd) = O(K^2d + Ndd')$.

\vspace{-0.05cm}
\subsection{Model Training and Inference }

We can describe the process of generating embeddings based on initial interactions as $X \to Z \to Y$, where $X$, $Z$, and $Y$ are random variables that respectively represent the initial interaction, the parameters of the personalized mapping function, and the user embedding. The goal is to learn the parameters of the personalized mapping function to maximize the expression of the ground user embedding and to compress the information contained in the initial interactions as much as possible by maximizing the information bottleneck objective\cite{tishby2000information}, as shown in the following equation:

\begin{equation}
\label{ib}
\mathcal{O} _{\text{IB}}= I(Z, Y)-\beta I(Z, X) 
\end{equation}

To optimize the objective function, we can use the variational information bottleneck (VIB) method\cite{vib}, which involves maximizing a lower bound on the information bottleneck objective. The lower bound is expressed as follows:

\begin{equation}
\begin{aligned}
 O_{\mathrm{IB}} \geq \mathbb{E}_u\left[\mathbb{E}_{\mathbf{w} \sim q(\mathbf{w} \mid I(u))}\left[\log p\left(\phi_u \mid \mathbf{w}\right)\right]\right.-\beta \mathrm{KL}[q(\mathbf{w} \mid I(u)), q(\mathbf{w})]]
\end{aligned}
\end{equation}
where $q(\mathbf{w}\mid I(u^c))$ is a variational spike-and-slab distribution used to approximate the true posterior distribution $p(\mathbf{w}\mid I(u^c))$, and $\beta$ is a hyperparameter.


The first term in the objective function indicates the desire for the log-likelihood to be maximized given the parameters $\mathbf{w}$ following the approximate distribution $q$. Thus, when the noise follows a Gaussian distribution, maximizing the log likelihood is equivalent to minimizing the mean squared error (MSE) between the predicted embedding $\hat{\phi}_{u}$ and the ground embedding $\phi_{u}$, which has already been pre-trained together with the recommender model.

The second term in the objective function indicates that the approximate distribution should have a small Kullback–Leibler (KL) divergence with the prior distribution. As the approximate distribution is a spike-and-slab distribution consisting of a Bernoulli part and a Gaussian part, we assume that the prior for the Bernoulli part is a Bernoulli distribution with parameter $0$, which encourages the model to learn sparse parameters, while the prior for the Gaussian part is a standard Gaussian distribution.

We employ the existing user, i.e. $\mathcal{U}^{e}$, as training data because we can obtain both their initial interactions and pre-trained embeddings as ground truth. Consequently, we are able to compute the loss function and optimize the model parameters $\Theta_{\text{PG}}$.

Considering the above, we minimize the final objective function as follows:
\begin{equation}
\label{loss}
\mathcal{L}=\mathcal{L} _{\text{MSE}}+\beta\mathcal{L} _{\text{KL}}
\end{equation}
\vspace{-0.2cm}
where



\vspace{-0.3cm}
\begin{equation}
\begin{aligned}
\mathcal{L}_{\text {MSE }}&=\frac{1}{\left|\mathcal{U}^e\right|} \sum_{i=1}^{\left|\mathcal{U}^e\right|}\left\|f\left(\Phi^U ; \mathbf{w}\right)-\phi_{u_i^e}\right\|_2 \\
\mathcal{L}_{\operatorname{KL}}&=\frac{1}{\left|\mathcal{U}^e\right|} \sum_{i=1}^{\left|\mathcal{U}^e\right|}[\operatorname{KL}(\text { Bern }(\boldsymbol{\pi}), \text { Bern }(\mathbf{0})) +\operatorname{KL}\left(\mathcal{N}\left(\boldsymbol{\mu}, \boldsymbol{\sigma}\right), \mathcal{N}(\mathbf{0}, \mathbf{I})\right)]
\end{aligned}
\end{equation}
\vspace{-0.2cm}

where $\beta$ controls the sparsity of the model, with higher values of $\beta$ leading to a sparser model.
It should be noted that the trainable parameters are only $\Theta_{\text{PG}}$, which includes the initial interaction encoder and the multi-layer perceptron used for reparameterization, while the parameters of the 
pre-trained base recommender model do not require fine-tuning.

During the inference phase, for a cold-start user $u^{c}$, a spike-and-slab distribution $\mathcal{S}(\boldsymbol{\pi}, \boldsymbol{\mu}, \boldsymbol{\sigma})$ is generated. Subsequently, parameters $\mathbf{w}$ are then sampled from this distribution, which serve as the parameters for the personalized mapping function $f$. Finally, the personalized mapping function is used to map the existing user embeddings $\Phi^U$ to the embeddings of this cold-start user $\hat{\phi}_{u^{c}}$, which is utilized to compute the click-through rate or ranking score. 


\section{Experiments}

We conduct extensive experiments to address the following research questions:

\textbf{RQ1}: How does VM-Rec perform compared to other cold-start methods?

\textbf{RQ2}: What is the impact of the hyperparameter on the performance?

\textbf{RQ3}: What is the impact of using different types of variational distributions?

\subsection{Experimental Settings }
\subsubsection{Datasets and Baselines}

We evaluate our method using MovieLens 100K\cite{harper2015movielens} (943 users and 1682 items), LastFM\cite{lastfm} (1892 users and 12523 items) and Ali Display Ad Click\cite{taobao} (491647 users and 240130 items) datasets. We selected these three datasets for two reasons: 1) they represent different scenarios, i.e., movie, music and e-commerce platforms, with different levels of item diversity, and 2) they differ in user and item scale.

We employ three well-known collaborative filtering models, namely bayesian personalized ranking (BPR)\cite{bpr}, neural graph collaborative filtering (NGCF)\cite{neuralgcf} and light graph convolution network (LightGCN)\cite{lightgcn}, as pre-trained base models prior to cold-start recommendation.
For each base model, we compare five baseline methods with our proposed VM-Rec as follows: 
1) Fine-tuning based on the pre-trained base recommender model with randomly initialized embeddings. 2) DropoutNet\cite{dropoutnet}, a method addresses the cold-start problem by utilizing neural network models with dropout and utilizing the average of content embeddings to serve as the embedding for a new ID. 3) MetaEmbedding\cite{metaemb}, a meta-learning-based method that generates desirable initial embeddings for new IDs, utilizing previously learned IDs through gradient-based meta-learning. 4) MWUF\cite{mwuf}, a recommendation framework, employs Meta Scaling and Shifting Networks to improve model fitting for cold ID embeddings (fast adaptation), alleviating noise influence and enhancing cold start recommendations. 5)WDoF\cite{wdof}, a method for user cold-start recommendation, models user preference as a weighted distribution over functions. It should be noted that WDoF is an autoencoder architecture model and does not include a base model.

\vspace{-0.12cm}
\subsubsection{Evaluation Protocol}


To simulate a cold-start scenario, we first sort the users in the dataset according to the time of their first interaction and then divide them into training, validation, and test sets in an 8:1:1 ratio. 
In this context, we utilize the training set to train the base recommender model, while the validation and test sets are utilized to evaluate the performance with respect to cold-start users. Importantly, the data partitioning scheme dictates that users within the training, validation, and test sets do not intersect, and furthermore, the user embeddings within the validation and test sets do not appear within the trained base recommender model.
Additionally, we only keep users who have two or more interactions to ensure that each user has at least one initial interaction item and one item for evaluation purposes. To mitigate the influence of cold-start items on evaluating cold-start user recommendations, items that appear in the validation and test sets but not in the training set are removed, because we cannot accurately calculate scores for a new item, and such items would not contribute to evaluate user cold-start methods. 

During the evaluation process, we employ a k-shot scenario to simulate user cold-start recommendations in the real-world context. For each user, we consider the first k (1, 2, or 3) items from their interaction sequences as observed initial interactions, while the remaining items are used to evaluate the performance of cold-start recommendations with a 1:100 positive-to-negative sample ratio.



\vspace{-0.12cm}
\subsubsection{Implementation Details}

\begin{table*}[!t]
\setlength\tabcolsep{1.0pt} 
\renewcommand{\arraystretch}{1.5}
\vspace{-0.5cm}
\caption{Experimental results of different methods.}
\label{tab:my-table}

\resizebox{\linewidth}{!}{

  \begin{tabular}{c|c|cccccc|cccccc|cccccc}
  \toprule
  \multirow{3}{*}{Model}           & \multirow{3}{*}{Base RS} & \multicolumn{6}{c|}{MovieLens-100K}                                                                                           & \multicolumn{6}{c|}{LastFM}                                                                                                   & \multicolumn{6}{c}{Ali Display Ad Click}                                                                                      \\ \cline{3-20} 
                                   &                          & \multicolumn{3}{c|}{NDCG@5}                                             & \multicolumn{3}{c|}{MRR@5}                          & \multicolumn{3}{c|}{NDCG@5}                                              & \multicolumn{3}{c|}{MRR@5}                         & \multicolumn{3}{c|}{NDCG@5}                                              & \multicolumn{3}{c}{MRR@5}                          \\ \cline{3-20} 
                                   &                          & 1-shot          & 2-shot          & \multicolumn{1}{c|}{3-shot}         & 1-shot          & 2-shot          & 3-shot          & 1-shot          & 2-shot          & \multicolumn{1}{c|}{3-shot}          & 1-shot         & 2-shot          & 3-shot          & 1-shot          & 2-shot          & \multicolumn{1}{c|}{3-shot}          & 1-shot          & 2-shot         & 3-shot          \\ \hline
  \multirow{3}{*}{Fine-tuning}     & BPR                      & 0.0284          & 0.0645          & \multicolumn{1}{c|}{0.0412}         & 0.0221          & 0.0516          & 0.0318          & 0.0588          & 0.0606          & \multicolumn{1}{c|}{0.0468}          & 0.0463         & 0.0471          & 0.0385          & 0.1             & 0.0963          & \multicolumn{1}{c|}{0.1031}          & 0.0851          & 0.0811         & 0.0874          \\
                                   & NGCF                     & 0.0656          & 0.057           & \multicolumn{1}{c|}{0.0736}         & 0.052           & 0.0468          & 0.0609          & 0.058           & 0.0416          & \multicolumn{1}{c|}{0.0547}          & 0.0458         & 0.0323          & 0.0446          & 0.0702          & 0.0679          & \multicolumn{1}{c|}{0.0682}          & 0.0584          & 0.0564         & 0.0567          \\
                                   & LightGCN                 & 0.0641          & 0.0511          & \multicolumn{1}{c|}{0.0458}         & 0.0524          & 0.0413          & 0.037           & 0.0606          & 0.0579          & \multicolumn{1}{c|}{0.0541}          & 0.0516         & 0.048           & 0.0453          & 0.1254          & 0.1205          & \multicolumn{1}{c|}{0.1149}          & 0.11            & 0.1055         & 0.1005          \\ \hline
  \multirow{3}{*}{DropoutNet}      & BPR                      & 0.1452          & 0.1552          & \multicolumn{1}{c|}{0.1589}         & 0.1206          & 0.1302          & 0.1299          & 0.3017          & 0.314           & \multicolumn{1}{c|}{0.314}           & 0.2679         & 0.2742          & 0.2752          & 0.1807          & 0.1975          & \multicolumn{1}{c|}{0.1995}          & 0.1512          & 0.164          & 0.1665          \\
                                   & NGCF                     & 0.108           & 0.1257          & \multicolumn{1}{c|}{0.1285}         & 0.0885          & 0.1052          & 0.1088          & 0.3156          & 0.3208          & \multicolumn{1}{c|}{0.3458}          & 0.2782         & 0.2794          & 0.3012          & 0.284           & 0.309           & \multicolumn{1}{c|}{0.3103}          & 0.2477          & 0.2697         & 0.2694          \\
                                   & LightGCN                 & 0.1301          & 0.1311          & \multicolumn{1}{c|}{0.14}           & 0.1075          & 0.1104          & 0.1179          & 0.2351          & 0.2259          & \multicolumn{1}{c|}{0.2211}          & 0.1988         & 0.1855          & 0.1807          & 0.1092          & 0.1015          & \multicolumn{1}{c|}{0.0953}          & 0.0923          & 0.0851         & 0.0794          \\ \hline
  \multirow{3}{*}{MetaEmbed}       & BPR                      & 0.2199          & 0.2313          & \multicolumn{1}{c|}{0.2341}         & 0.1831          & 0.1946          & 0.1958          & 0.2619          & 0.2585          & \multicolumn{1}{c|}{0.2587}          & 0.2246         & 0.2221          & 0.222           & 0.3515          & 0.3616          & \multicolumn{1}{c|}{0.3844}          & 0.3143          & 0.3233         & 0.3444          \\
                                   & NGCF                     & 0.2663          & 0.277           & \multicolumn{1}{c|}{0.2773}         & 0.2246          & 0.2324          & 0.2324          & 0.253           & 0.2462          & \multicolumn{1}{c|}{0.2489}          & 0.2187         & 0.2133          & 0.2155          & 0.3116          & 0.3194          & \multicolumn{1}{c|}{0.3413}          & 0.2773          & 0.2839         & 0.3037          \\
                                   & LightGCN                 & 0.2707          & 0.2742          & \multicolumn{1}{c|}{0.2762}         & 0.229           & 0.232           & 0.2323          & 0.2964          & 0.2912          & \multicolumn{1}{c|}{0.297}           & 0.2596         & 0.2545          & 0.2584          & 0.2659          & 0.2735          & \multicolumn{1}{c|}{0.291}           & 0.2376          & 0.2437         & 0.2602          \\ \hline
  \multirow{3}{*}{MWUF}            & BPR                      & 0.2591          & 0.2563          & \multicolumn{1}{c|}{0.2534}         & 0.2164          & 0.213           & 0.21            & 0.2952          & 0.3492          & \multicolumn{1}{c|}{0.3817}          & 0.2551         & 0.3047          & 0.3347          & 0.3625          & 0.4326          & \multicolumn{1}{c|}{0.5294}          & 0.3253          & 0.3936         & 0.4901          \\
                                   & NGCF                     & 0.2619          & 0.2658          & \multicolumn{1}{c|}{0.2655}         & 0.2192          & 0.2208          & 0.2204          & 0.2795          & 0.344           & \multicolumn{1}{c|}{0.3675}          & 0.2421         & 0.3031          & 0.3245          & 0.3092          & 0.4834          & \multicolumn{1}{c|}{0.5873}          & 0.2706          & 0.4428         & 0.5443          \\
                                   & LightGCN                 & 0.2411          & 0.2374          & \multicolumn{1}{c|}{0.2256}         & 0.1998          & 0.1978          & 0.1893          & 0.3102          & 0.3677          & \multicolumn{1}{c|}{0.3906}          & 0.2743         & 0.3281          & 0.3468          & 0.2598          & 0.4869          & \multicolumn{1}{c|}{0.5883}          & 0.2283          & 0.4456         & 0.5465          \\ \hline
  WDoF                             & -                        & 0.1651          & 0.2216          & \multicolumn{1}{c|}{0.2831}         & 0.1388          & 0.1854          & 0.2366          & 0.234           & 0.2638          & \multicolumn{1}{c|}{0.289}           & 0.2011         & 0.2308          & 0.2476          & 0.2819          & 0.312           & \multicolumn{1}{c|}{0.3326}          & 0.2507          & 0.2779         & 0.2952          \\ \hline
  \multirow{3}{*}{\textbf{VM-Rec}} & BPR                      & \underline{ 0.2851}    & 0.3035          & \multicolumn{1}{c|}{0.3098}         & 0.2425          & 0.2579          & 0.2651          & \underline{ 0.4073}    &  0.4322    & \multicolumn{1}{c|}{0.4524}          & \underline{ 0.3654}   & 0.3898          & 0.409           & 0.4105          & 0.5048          & \multicolumn{1}{c|}{0.5186}          & 0.3742          & 0.4652         & 0.4768          \\
                                   & NGCF                     & 0.2839          & \textbf{0.3045} & \multicolumn{1}{c|}{\underline{ 0.3102}}   & \underline{ 0.2433}    & \textbf{0.2616} & \textbf{0.2675} & \textbf{0.4156} & \underline{0.4408}          & \multicolumn{1}{c|}{\underline{ 0.4563}}    & \textbf{0.372} & \underline{ 0.398}     & \underline{ 0.4108}    & \underline{ 0.4841}    & \textbf{0.6209} & \multicolumn{1}{c|}{\textbf{0.6422}} & \underline{ 0.446}     & \textbf{0.584} & \textbf{0.6013} \\
                                   & LightGCN                 & \textbf{0.2865} & \underline{ 0.3038}    & \multicolumn{1}{c|}{\textbf{0.311}} & \textbf{0.2449} & \underline{ 0.259}     & \underline{ 0.2665}    & 0.3939          & \textbf{0.4445} & \multicolumn{1}{c|}{\textbf{0.4629}} & 0.3544         & \textbf{0.4028} & \textbf{0.4216} & \textbf{0.5086} & \underline{ 0.6188}    & \multicolumn{1}{c|}{\underline{ 0.633}}     & \textbf{0.4719} & \underline{ 0.5795}   & \underline{ 0.5922}    \\ \bottomrule
  \end{tabular}

}
\vspace{-0.4cm}
\end{table*}

We implement the base models using RecBole\cite{recbole[1.1.1]}, setting the embedding dimension to 64 and the learning rate to 0.001 uniformly, while utilizing the default configurations for other hyperparameters in RecBole.

When training the cold-start model, the following procedures are adopted for each method: In the case of fine-tuning, a random embedding is generated to update using the observed k initial interactions. For DropoutNet and MWUF, user/item historical interactions are treated as user/item features, following \cite{metacf}. In the case of MetaEmbedding, the generator produces a global initial embedding for cold-start users, which is subsequently updated using the initial interactions.
For DropoutNet\numfootnote{1}{\url{https://github.com/layer6ai-labs/DropoutNet}} and WDoF\numfootnote{2}{\url{https://github.com/xuan92ta/WDoF}}, the settings follow the official implementation.
For VM-Rec, the number of heads in the multi-head self-attention used in the initial interactions encoder is set to 4, the hidden layer dimension is set to 64, $\operatorname{MLP}^{(1)}_{\pi/\mu/\sigma}$ is set to $(128, 128)$, $\operatorname{MLP}^{(2)}_{\pi}$  is set to $(128, 1)$, $\operatorname{MLP}^{(2)}_{\mu/\sigma}$ is set to $(256, 1)$, and hyperparameter $\beta$ is set to $10^{-10}$. When training VM-Rec on the Ali Display Ad Click dataset, sampled 50,000 existing user embeddings were used considering computational efficiency. In section \ref{proportions}, we will discuss the impact of using embeddings with different proportions on performance.
During the training, the weights of the base recommender model are frozen. The Adam optimizer is employed, with the learning rate searched from \{0.1, 0.01, 0.001, 0.0001\}, and weight decay is set to $10^{-5}$. To mitigate overfitting, the maximum number of training epochs is restricted to 50, and early stopping is implemented with a patience of 3 based on the performance on the validation set. Ultimately, the NDCG@5 and MRR@5 are applied to the test set for evaluation purposes.

\subsection{Experimental Results}
\subsubsection{Overall Performance (RQ1)}

We implemented 6 cold-start methods based on 3 base recommender models and compared their performances on 3 datasets, as illustrated in Table \ref{tab:my-table}.
Based on the experimental results, we have the following three observations:

\textbf{Observation 1:} VM-Rec outperforms other selected cold-start methods on three datasets, demonstrating its ability to adapt to different datasets of varying scales and scenarios.

\textbf{Observation 2:} The performance of VM-Rec in 1, 2, and 3-shot scenarios shows an increasing trend, indicating that the model benefits positively from an increase in initial interactions. 
However, content-based approaches such as DropoutNet and MWUF face a substantial gap in content features between training and few-shot inference due to artificially construct content features, followed \cite{metacf}.
When updating embeddings based on random initialization or learned meta embeddings, the inherent stochasticity in few-shot scenarios leads to unstable update directions, and the performance does not consistently improve with an increase in shots as well. WDoF illustrates an increase in support size (shot), but in Ali dataset, characterized by high item diversity in the e-commerce context, modeling user preferences becomes challenging, thereby impacting performance.

\begin{figure*}[t]
    \centering
        \begin{subfigure}{0.27\textwidth}
            \centering
            \includegraphics[width=\linewidth]{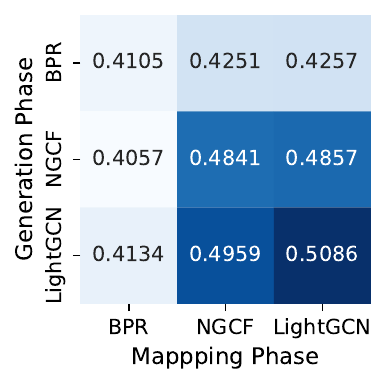}
            \caption{NDCG@5 using different base models.}
            \label{fig:cross}
        \end{subfigure}
        \hfill
        \begin{subfigure}{0.33\textwidth}
            \centering
            \includegraphics[width=\linewidth]{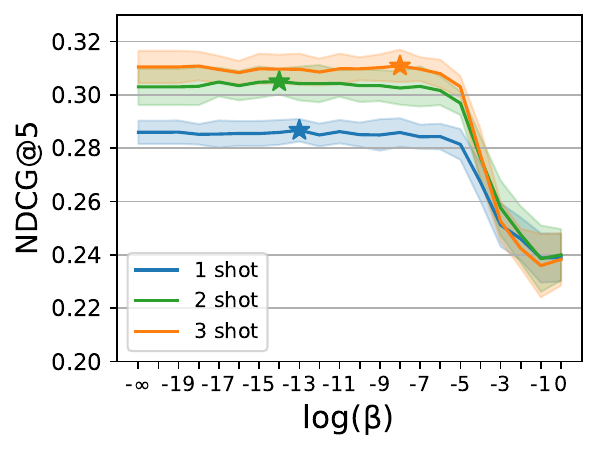}
            \caption{NDCG@5 across different values of $\beta$.}
            \label{fig:beta}
        \end{subfigure}
        \hfill
        \begin{subfigure}{0.335\textwidth}
            \centering
            \includegraphics[width=\linewidth]{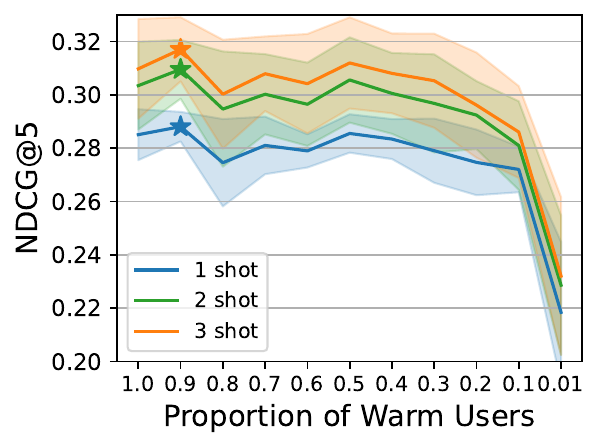}
            \caption{NDCG@5 across different proportions.}
            \label{fig:prop}
        \end{subfigure}

    \caption{The impact of varying base model, $\beta$ and proportion on performance.}
\end{figure*}

\textbf{Observation 3:} VM-Rec, when based on BPR as the base model, generally performs worse than NGCF and LightGCN. The phenomenon is especially notable on larger datasets like Ali. This suggests that the quality of embeddings in VM-Rec, which relies on embedding mapping, influences the effectiveness of cold-start recommendations. 
As depicted in Figure \ref{fig-model}, all trainable parameters are situated within the parameter generation phase.  Consequently, we explored employing different base models for the two phases, leading to an intriguing discovery.
Despite the suboptimal performance of VM-Rec trained with BPR on the Ali dataset, transferring it to NGCF and LightGCN for mapping results in performance improvement, as shown in Figure \ref{fig:cross}. This indicates that 1) VM-Rec's learned strategies for selecting warm instances and weighting them are reasonable, and 
2) This characteristic of VM-Rec holds promise for applications in train-free cross-model deployment or cross-domain recommendation scenarios 
with user overlap.


\vspace{-0.5cm}

\vspace{-0.12cm}
\subsubsection{Sensitivity Analysis of Hyperparameters (RQ2)}\label{proportions}

For the hyperparameter $\beta$ in the objective Equation \ref{loss}, as shown in Figure \ref{fig:beta}. We have observed that selecting an appropriate value for $\beta$ is advantageous for enhancing the model's performance. 
when $\beta$ is small, inadequate regularization leads to a reduced generalization
capacity of the model, thereby affecting its performance.
Conversely, an excessively large $\beta$ will encourage the mapping function to be overly sparse, rendering it ineffective for accurate predictions.


For using different proportions of warm user embeddings, as shown in Figure \ref{fig:prop}, the performance will significantly decrease when using a too small proportion and when the proportion is 0.9, it is the optimal choice for 1, 2, and 3-shot scenarios. 
This phenomenon suggests that not training with all warm user embeddings may not necessarily result in a deterioration of performance. The inclusion of too many users poses challenges in masking irrelevant users and employing a subset of users is deemed acceptable, unless the proportion is extremely small.

\begin{table}[t]
\begin{minipage}{0.48\textwidth}
\small
\centering
\caption{NDCG@5 using different distributions.}
\label{tab-dist}
\begin{tabular}{@{}cccc@{}}
\toprule
Distribution   & 1-shot & 2-shot & 3-shot \\ \midrule
Gaussian       & 0.0982 & 0.0954 & 0.0883 \\
Gaussian+L1    & 0.1015 & 0.1023 & 0.0955 \\
Spike-and-Slab & 0.2851 & 0.3035 & 0.3098 \\ \bottomrule
\end{tabular}
\end{minipage}%
\hfill
\begin{minipage}{0.48\textwidth}
\centering
\small
\vspace{-0.35cm}
\caption{NDCG@5 using different mapping types.}
\label{tab-rm}
\begin{tabular}{@{}ccccc@{}}
\toprule
Model    & Subset & 1-shot & 2-shot & 3-shot \\ \midrule
RM\_init & easy   & 0.2799 & 0.3019 & 0.4047 \\
RM\_cont & easy   & 0.2784 & 0.298  & 0.4047 \\
VM-Rec       & easy   & 0.3021 & 0.3299 & 0.3828 \\
VM-Rec       & hard  & 0.2019 & 0.2586 & 0.3017 \\ \bottomrule
\end{tabular}
\end{minipage}
\end{table}

\renewcommand{\thefootnote}{}
\footnotetext{Figure \ref{fig:cross} is conducted on Ali dataset for 1 shot. Figure \ref{fig:beta}, \ref{fig:prop} and Table \ref{tab-dist}, \ref{tab-rm} are conducted on Movielens dataset using BPR as base model.}

\vspace{-0.12cm}
\subsubsection{Ablation Experiment on the Type of Variational Distribution (RQ3)} \label{Ablation}

In order to investigate the impact of different variational distributions on performance, we compared the spike-and-slab distribution, the Gaussian distribution, and the Gaussian distribution with L1 regularization applied to the mapping function's parameters $\mathbf{w}$. The regularization coefficient was searched from $\left\{10^{-20}, \dots, 1, 10, 100\right\}$. 
As shown in Table \ref{tab-dist}, when using Gaussian with L1 regularization, the performance decreases by an average of 68.54\% and 66.62\%, respectively, across 1, 2, and 3 shot scenarios.
When ignoring sparsity and solely using Gaussian distribution to generate parameters, the excessive number of warm user embeddings becomes problematic as not all of them contribute significantly to predictions.
 L1 regularization can achieve some sparsity, but it penalizes all weights uniformly. In contrast, the spike-and-slab distribution is considered the best option for sparse linear models. It offers a way to achieve both sparsity and model flexibility effectively\cite{titsias2011spike}. This distribution allows for precise control over the probability of each parameter being \textit{sparse}.


In addition, we also investigated two rule-based mappings (RM) that are similarly sparse. 1) RM\_init averages the embeddings of existing warm users who have the same initial interactions as cold-start users. 2) RM\_cont averages the embeddings of existing warm users with continuous historical interactions same as cold-start users' initial interactions. We further partition the users in the test set into two non-overlapping subsets, denoted as $\mathcal{U}^{\text{easy}}$ and $\mathcal{U}^{\text{hard}}$. For $u^c \in \mathcal{U}^{\text{easy}}$, we require that
$I(u^c)\in \{I(u^e)|u^e \in \mathcal{U}^e\}$. $\mathcal{U}^{\text{hard}}$ follows the opposite condition.
For the Movielens dataset in 1, 2, and 3 shot scenarios, $\mathcal{U}^{\text{easy}}$ accounts for 83.15\%, 48.42\%, and 4.21\%, respectively.
The performance of rule-based mapping (RM) and variational mapping on different subsets is shown in Table \ref{tab-rm}. It can be observed that, for $\mathcal{U}^{\text{easy}}$, VM-Rec outperforms both types of RM, and RM\_init is comparable to RM\_cont. This suggests that, when predicting embeddings, warm users with same initial interactions are more meaningfully associated than those with continuous interactions in the history. We also found that $\mathcal{U}^{\text{easy}}$ is superior to $\mathcal{U}^{\text{hard}}$ because the initial interactions in $\mathcal{U}^{\text{easy}}$ have been encountered during training, making them easier to model, while the initial interactions in $\mathcal{U}^{\text{hard}}$ test the generalization ability of VM-Rec. However, rule-based methods such as RM\_init and RM\_cont cannot handle the cold-start recommendation for $\mathcal{U}^{\text{hard}}$. Through ablation experiments, we can verify the assumption that initial interactions to some extent reflect the distribution of embeddings and confirm the superiority of VM-Rec in the strategy of selecting and weighting warm user embeddings.



\vspace{-0.1cm}
\section{CONCLUSIONS}

We investigates the user cold-start problem in recommender systems and propose a variational mapping approach that
generates a personalized mapping function for cold-start users to map the embeddings of existing users into a new embedding. We then evaluate our approach
on two datasets and three base recommender models, showing that our proposed
VM-Rec outperforms baseline methods. Finally, sensitivity analysis and ablation
experiments demonstrate the impact of hyperparameters and variational distributions on performance.

\vspace{-0.1cm}
%
%
%
%

\bibliographystyle{splncs04}
\bibliography{sample-base}

\begin{thebibliography}{10}
\providecommand{\url}[1]{\texttt{#1}}
\providecommand{\urlprefix}{URL }
\providecommand{\doi}[1]{https://doi.org/#1}

\bibitem{vib}
Alemi, A.A., Fischer, I., Dillon, J.V., Murphy, K.: Deep variational information bottleneck. arXiv preprint arXiv:1612.00410  (2016)

\bibitem{lastfm}
Cantador, I., Brusilovsky, P., Kuflik, T.: Second workshop on information heterogeneity and fusion in recommender systems (hetrec2011). In: Proceedings of the fifth ACM conference on Recommender systems. pp. 387--388 (2011)

\bibitem{chen2022generative}
Chen, H., Wang, Z., Huang, F., Huang, X., Xu, Y., Lin, Y., He, P., Li, Z.: Generative adversarial framework for cold-start item recommendation. In: Proceedings of the 45th International ACM SIGIR Conference on Research and Development in Information Retrieval. pp. 2565--2571 (2022)

\bibitem{dai2021poso}
Dai, S., Lin, H., Zhao, Z., Lin, J., Wu, H., Wang, Z., Yang, S., Liu, J.: Poso: Personalized cold start modules for large-scale recommender systems. arXiv preprint arXiv:2108.04690  (2021)

\bibitem{finn2017model}
Finn, C., Abbeel, P., Levine, S.: Model-agnostic meta-learning for fast adaptation of deep networks. In: International conference on machine learning. pp. 1126--1135. PMLR (2017)

\bibitem{gan}
Goodfellow, I., Pouget-Abadie, J., Mirza, M., Xu, B., Warde-Farley, D., Ozair, S., Courville, A., Bengio, Y.: Generative adversarial networks. Communications of the ACM  \textbf{63}(11),  139--144 (2020)

\bibitem{guo2016entity}
Guo, C., Berkhahn, F.: Entity embeddings of categorical variables. arXiv preprint arXiv:1604.06737  (2016)

\bibitem{hao2021pre}
Hao, B., Zhang, J., Yin, H., Li, C., Chen, H.: Pre-training graph neural networks for cold-start users and items representation. In: Proceedings of the 14th ACM International Conference on Web Search and Data Mining. pp. 265--273 (2021)

\bibitem{harper2015movielens}
Harper, F.M., Konstan, J.A.: The movielens datasets: History and context. Acm transactions on interactive intelligent systems (tiis)  \textbf{5}(4),  1--19 (2015)

\bibitem{lightgcn}
He, X., Deng, K., Wang, X., Li, Y., Zhang, Y., Wang, M.: Lightgcn: Simplifying and powering graph convolution network for recommendation. In: Proceedings of the 43rd International ACM SIGIR conference on research and development in Information Retrieval. pp. 639--648 (2020)

\bibitem{gumbel-softmax}
Jang, E., Gu, S., Poole, B.: Categorical reparameterization with gumbel-softmax. arXiv preprint arXiv:1611.01144  (2016)

\bibitem{kim2023meta}
Kim, M., Yang, Y., Ryu, J.H., Kim, T.: Meta-learning with adaptive weighted loss for imbalanced cold-start recommendation. arXiv preprint arXiv:2302.14640  (2023)

\bibitem{kingma2013auto}
Kingma, D.P., Welling, M.: Auto-encoding variational bayes. arXiv preprint arXiv:1312.6114  (2013)

\bibitem{lee2019melu}
Lee, H., Im, J., Jang, S., Cho, H., Chung, S.: Melu: Meta-learned user preference estimator for cold-start recommendation. In: Proceedings of the 25th ACM SIGKDD International Conference on Knowledge Discovery \& Data Mining. pp. 1073--1082 (2019)

\bibitem{liu2023Boosting}
Liu, H., Lin, H., Zhang, X., Ma, F., Chen, H., Wang, L., Yu, H., Zhang, X.: Boosting meta-learning cold-start recommendation with graph neural network. In: Proceedings of the 32nd ACM International Conference on Information and Knowledge Management. p. 4105–4109. CIKM '23 (2023)

\bibitem{lu2020meta}
Lu, Y., Fang, Y., Shi, C.: Meta-learning on heterogeneous information networks for cold-start recommendation. In: Proceedings of the 26th ACM SIGKDD International Conference on Knowledge Discovery \& Data Mining. pp. 1563--1573 (2020)

\bibitem{van2008visualizing}
Van~der Maaten, L., Hinton, G.: Visualizing data using t-sne. Journal of machine learning research  \textbf{9}(11) (2008)

\bibitem{sas}
Mitchell, T.J., Beauchamp, J.J.: Bayesian variable selection in linear regression. Journal of the american statistical association  \textbf{83}(404),  1023--1032 (1988)

\bibitem{ouyang2021learning}
Ouyang, W., Zhang, X., Ren, S., Li, L., Zhang, K., Luo, J., Liu, Z., Du, Y.: Learning graph meta embeddings for cold-start ads in click-through rate prediction. In: Proceedings of the 44th International ACM SIGIR Conference on Research and Development in Information Retrieval. pp. 1157--1166 (2021)

\bibitem{metaemb}
Pan, F., Li, S., Ao, X., Tang, P., He, Q.: Warm up cold-start advertisements: Improving ctr predictions via learning to learn id embeddings. In: Proceedings of the 42nd International ACM SIGIR Conference on Research and Development in Information Retrieval. pp. 695--704 (2019)

\bibitem{bpr}
Rendle, S., Freudenthaler, C., Gantner, Z., Schmidt-Thieme, L.: Bpr: Bayesian personalized ranking from implicit feedback. arXiv preprint arXiv:1205.2618  (2012)

\bibitem{taobao}
Tianchi: Ad display/click data on taobao.com (2018), \url{https://tianchi.aliyun.com/dataset/dataDetail?dataId=56}

\bibitem{tishby2000information}
Tishby, N., Pereira, F.C., Bialek, W.: The information bottleneck method. arXiv preprint physics/0004057  (2000)

\bibitem{titsias2011spike}
Titsias, M., L{\'a}zaro-Gredilla, M.: Spike and slab variational inference for multi-task and multiple kernel learning. Advances in neural information processing systems  \textbf{24} (2011)

\bibitem{vaswani2017attention}
Vaswani, A., Shazeer, N., Parmar, N., Uszkoreit, J., Jones, L., Gomez, A.N., Kaiser, {\L}., Polosukhin, I.: Attention is all you need. Advances in neural information processing systems  \textbf{30} (2017)

\bibitem{dropoutnet}
Volkovs, M., Yu, G., Poutanen, T.: Dropoutnet: Addressing cold start in recommender systems. Advances in neural information processing systems  \textbf{30} (2017)

\bibitem{neuralgcf}
Wang, X., He, X., Wang, M., Feng, F., Chua, T.S.: Neural graph collaborative filtering. In: Proceedings of the 42nd international ACM SIGIR conference on Research and development in Information Retrieval. pp. 165--174 (2019)

\bibitem{metacf}
Wei, T., Wu, Z., Li, R., Hu, Z., Feng, F., He, X., Sun, Y., Wang, W.: Fast adaptation for cold-start collaborative filtering with meta-learning. In: 2020 IEEE International Conference on Data Mining (ICDM). pp. 661--670. IEEE (2020)

\bibitem{wei2021contrastive}
Wei, Y., Wang, X., Li, Q., Nie, L., Li, Y., Li, X., Chua, T.S.: Contrastive learning for cold-start recommendation. In: Proceedings of the 29th ACM International Conference on Multimedia. pp. 5382--5390 (2021)

\bibitem{wdof}
Wen, J., Liu, H., Jing, L.: Modeling preference as weighted distribution over functions for user cold-start recommendation. In: Proceedings of the 32nd ACM International Conference on Information and Knowledge Management. p. 2706–2715. CIKM '23 (2023)

\bibitem{recbole[1.1.1]}
Xu, L., Tian, Z., Zhang, G., Wang, L., Zhang, J., Zheng, B., Li, Y., Hou, Y., Pan, X., Chen, Y., Zhao, W.X., Chen, X., Wen, J.R.: Recent advances in recbole: Extensions with more practical considerations (2022)

\bibitem{velf}
Xu, X., Yang, C., Yu, Q., Fang, Z., Wang, J., Fan, C., He, Y., Peng, C., Lin, Z., Shao, J.: Alleviating cold-start problem in ctr prediction with a variational embedding learning framework. In: Proceedings of the ACM Web Conference 2022. pp. 27--35 (2022)

\bibitem{yuan2020parameter}
Yuan, F., He, X., Karatzoglou, A., Zhang, L.: Parameter-efficient transfer from sequential behaviors for user modeling and recommendation. In: Proceedings of the 43rd International ACM SIGIR conference on research and development in Information Retrieval. pp. 1469--1478 (2020)

\bibitem{yuan2023go}
Yuan, Z., Yuan, F., Song, Y., Li, Y., Fu, J., Yang, F., Pan, Y., Ni, Y.: Where to go next for recommender systems? id-vs. modality-based recommender models revisited. arXiv preprint arXiv:2303.13835  (2023)

\bibitem{zhao2018learning}
Zhao, K., Li, Y., Shuai, Z., Yang, C.: Learning and transferring ids representation in e-commerce. In: Proceedings of the 24th ACM SIGKDD International Conference on Knowledge Discovery \& Data Mining. pp. 1031--1039 (2018)

\bibitem{zhao2022improving}
Zhao, X., Ren, Y., Du, Y., Zhang, S., Wang, N.: Improving item cold-start recommendation via model-agnostic conditional variational autoencoder. In: Proceedings of the 45th International ACM SIGIR Conference on Research and Development in Information Retrieval. pp. 2595--2600 (2022)

\bibitem{zhou2023contrastive}
Zhou, Z., Zhang, L., Yang, N.: Contrastive collaborative filtering for cold-start item recommendation. arXiv preprint arXiv:2302.02151  (2023)

\bibitem{mwuf}
Zhu, Y., Xie, R., Zhuang, F., Ge, K., Sun, Y., Zhang, X., Lin, L., Cao, J.: Learning to warm up cold item embeddings for cold-start recommendation with meta scaling and shifting networks. In: Proceedings of the 44th International ACM SIGIR Conference on Research and Development in Information Retrieval. pp. 1167--1176 (2021)

\bibitem{zhu2020recommendation}
Zhu, Z., Sefati, S., Saadatpanah, P., Caverlee, J.: Recommendation for new users and new items via randomized training and mixture-of-experts transformation. In: Proceedings of the 43rd International ACM SIGIR Conference on Research and Development in Information Retrieval. pp. 1121--1130 (2020)

\end{thebibliography}

\end{document}